# Fluorescence resonance energy transfer between organic dyes adsorbed onto nano-clay and Langmuir-Blodgett (LB) films

Syed Arshad Hussain[1,2,*], S. Chakraborty[1], D. Bhattacharjee[1] and R. A. Schoonheydt[2]

1. Department of Physics, Tripura University, Suryamaninagar-799130, Tripura, India.
2. Centre for Surface Chemistry and Catalysis, K.U.Leuven, Kasteelpark Arenberg 23, 3001 Leuven, Belgium

* Corresponding author
Email: sa_h153@hotmail.com, tuphysic@sancharnet.in (S. A. Hussain)
Phone: +913812375317 (O)
Fax: +913812374802 (O)

**Abstract:** In this communication we investigate two dyes $N,N'$-dioctadecyl thiacyanine perchlorate (NK) and octadecyl rhodamine B chloride (RhB) in Langmuir and Langmuir-Blodgett (LB) films with or with out a synthetic clay laponite. Observed changes in isotherms of RhB in absence and presence of nano-clay platelets indicate the incorporation of clay platelets onto RhB-clay hybrid films. AFM image confirms the incorporation of clay in hybrid films. FRET was observed in clay dispersion and LB films with and without clay. Efficiency of energy transfer was maximum in LB films with clay.

**Keywords:** Fluorescence Resonance Energy Transfer (FRET), Langmuir-Blodgett, pressure-area isotherm, clay, dyes, Atomic Force Microscope (AFM).

## 1. Introduction

Fluorescence Resonance Energy Transfer (FRET) is a physical phenomenon first described over 50 years ago [1,2]. Due to its sensitivity to distance, FRET has been used to investigate molecular level interaction. Fluorescence emission rate of energy transfer has wide applications in biomedical, protein folding, RNA/DNA identification and their energy transfer process [3-9]. Another important application of energy transfer is in dye lasers. Dye lasers have some limitations as the dye solution used as an active medium absorbs energy from the excitation source in a very limited spectral range and so the emission band also has these limitations. If a dye laser has to be used as an ideal source its spectral range needs to be extended. In order to extend the spectral range of operation, mixtures of different dye solutions/dye molecules embedded in solid matrices are being used. The work on energy transfer between different dye molecules in such mixtures in various solvents and solid matrices is, therefore, of great importance. The use of such energy transfer in dye lasers is also helpful in minimizing the photo-quenching effects and thereby, increasing the laser efficiency.

FRET is the relaxation of an excited donor molecule $D$ by transfer of its excited energy to an acceptor molecule $A$ [1,2]. Consequently, the later emits a photon. The non-radiative energy transfer occurs as a result of dipole-dipole coupling between the donor and the acceptor, and does not involve the emission and reabsorption of photons. The process can be represented as

$$D + h\upsilon \rightarrow D^*$$

$$D^* + A \rightarrow D + A^*$$

$$A^* \rightarrow A + h\upsilon$$

[Where $h$ is the Planck's constant and $\upsilon$ is the frequency of the radiation]

The two main conditions for FRET to occur are (i) sufficient overlap between the absorption band of acceptor ($A$) and the fluorescence band of donor ($D$) and (ii) the molecule $A$ must be in the neighborhood of the molecules $D$, as the efficiency of FRET depends on the inverse sixth power of the distance of separation between the donor and acceptor. Considering the proximity of the donor and acceptor molecules, FRET gives an opportunity to investigate and estimate distances between the two molecules. The relative orientation of the donor and acceptor transition dipoles also effect the FRET process.

FRET is effective over distance ranging in between 1 to 10 nm. The intervening of solvent or other macromolecules has little effect on the efficiency of the energy transfer, which depends primarily on the donor-acceptor distance [1,2].

Clay mineral particles play an important role in concentrating the dye molecules onto their surfaces. This may provide a platform for close interaction between energy donor and acceptor to come into close interaction making energy transfer possible, in contrast to inactive systems based on homogeneous solutions. There are very few reports on the energy transfer occurring in systems composed of organic dyes and clays [10-15]. Probably the first reports of efficient FRET among dyes adsorbed onto clay mineral sheets, are clay mineral dispersions with cyanine (energy donor) and rhodamine (energy acceptor) dyes simultaneously adsorbed onto clay mineral surfaces [14,15]. So it is extremely important to identify suitable energy donor-acceptor pair and to probe the condition for efficient energy transfer.

On the other hand Langmuir-Blodgett (LB) technique is a conventional and highly sophisticated method for fabricating organic ultrathin films, which are made by stepwise deposition of monolayer's of amphiphile from the air-water interface onto solid substrate [16]. With respect to other film preparation techniques, the LB technique has the added advantage of having smooth, uniform and almost defect free monolayer film [16].

In our present study, the trace of energy transfer between two dyes has been reported. Here, we used two cationic dyes namely, a thiacyanine dye 3-octadecyl-2-[(3-octadecyl-2(3H)-benzolylidene) methyl] benzothiazolium perchlorate or $N, N'$-dioctadecyl thiacyanine perchlorate (NK) and a xanthane dye octadecyl rhodamine B chloride (RhB) representing a suitable energy donor and acceptor respectively. We investigated the system in chloroform solutions, clay dispersions and organized in the Langmuir and Langmuir-Blodgett (LB) monolayer with and without a synthetic clay laponite.

These two dyes NK and RhB are in principle suitable for FRET. Both the dyes are highly fluorescent. The fluorescence spectrum of NK sufficiently overlaps with the absorption spectrum of RhB. The aim of this study was to investigate the feasibility of FRET in between NK and RhB adsorbed in smectites as well as also onto LB monolayer.

## 2. Experimental
### 2.1. Materials

NK [Hayashibara Biochemical Laboratories Inc] and RhB [Molecular Probes] were used as received. Molecular structures of the dyes are shown in the inset of figure 1. The dyes used in our studies are positively charged and the laponite particles are negatively charged. The dyes were dissolved in either HPLC grade chloroform (99.9 % Aldrich, stabilized by 0.5-1% ethanol) or HPLC grade methanol [Acros Organics, USA]. The clay mineral used in the present work was Laponite, obtained from Laponite Inorganics, UK and used as received. The size of the clay platelet was less than 0.05 $\mu m$ and CEC is 0.739 meq/gm. The clay dispersion used was of 2 ppm of laponite stirred for 24 hours in Milli-Q ultrapure water (electrical resistivity 18.2 M$\Omega$ cm$^{-1}$) with a magnetic stirrer.

### 2.2. Isotherm measurement and LB film preparation

A commercially available Langmuir-Blodgett (LB) film deposition instrument (Apex-2000C, India) was used for isotherm measurement and LB film preparation. Dilute solutions (30 $\mu l$ (conc=$10^{-3} M$) of RhB, NK or RhB + NK) was spread onto either Milli Q (electrical resistivity 18.2 M$\Omega$ cm$^{-1}$) water or clay dispersion in the LB trough. A surface pressure-area per molecule isotherm was obtained by a continuous compression of the dye monolayer at the air-water or air-clay dispersion interface of the LB trough by a barrier system. The surface pressure at the interface was measured using wilhelmy plate arrangement attached to a microbalance, whose output was interfaced to a microcomputer, which controlled the barrier movement confining the monolayer at the air-water interface. Either Milli-Q water or clay dispersion (2 ppm) was used as subphase and the temperature was maintained at $24^0$C.

Before each isotherm measurement, the trough and barrier were cleaned with ethanol and then rinsed by Milli-Q water. In addition the glass ware was thoroughly cleaned prior to the use. The surface pressure fluctuation was estimated to be less than 0.5 mN/m during the compression of the entire trough surface area. Then the barrier was moved back to its initial position and the sample containing monolayer forming material was spread on the subphase using a microsyringe. After a delay of 30 minutes, to evaporate the solvent, the film at the air water interface was compressed slowly at the rate of 5 mm/min to obtain a single surface pressure versus area per molecule ($\pi$-A) isotherm. All isotherms were run several times with freshly prepared solution.

Monolayer LB films were deposited in upstroke (lifting speed 5 mm / min) at a fixed surface pressure of 15 mN/m onto fluorescence grade quartz substrate.

### 2.3. UV-Vis Absorption and Fluorescence spectra measurement

UV-Vis absorption and fluorescence spectra of the solutions and films were recoded by a Perkin Elmer Lambda-25 Spectrophotometer and Perkin Elmer LS-55 Fluorescence Spectrophotometer respectively. For absorption measurement the films were kept perpendicular to the incident light and a clean quartz slide was used as reference. The fluorescence light was collected from the sample surface at an angle of $45^0$ (front face geometry) and the excitation wavelength ($\lambda_{ex}$) was 430 nm.

### 3. Background of Fluorescence Resonance Energy Transfer (FRET)

From a quantitative point of view, a transfer efficiency E can be defined as the fraction of the photons adsorbed by the donor that are transferred to the acceptor. According to the Förster model for energy transfer [17-19], E is governed by the critical radius (Förster distance) $R_0$, it is the donor-acceptor distance at which the transfer rate $K_F(r)$ is equal to the decay rate ($\tau_D$) of the donor in the absence of energy transfer according to the relation:

$$K_F(r) = \frac{1}{\tau_D}(\frac{R_0}{r})^6 \qquad (1)$$

$r$ is the distance between energy donor and acceptor.

$$\text{and } E = \frac{K_F}{\tau_D^{-1} + K_F} = \frac{\tau_D K_F}{1 + \tau_D K_F}, \qquad (2)$$

which is the ratio of the transfer rate to the total decay rate of the donor. Using the above two relations expressions for E can be rearranged to yield

$$E = \frac{R_0^6}{R_0^6 + r^6} \qquad (3)$$

In the present case the efficiency of the energy transfer (E) was calculated as $E = 1 - \frac{F_{DA}}{F_D}$, where $F_{DA}$ is the fluorescence intensity of the donor in the presence of acceptor, $F_D$ is the fluorescence intensity of the donor in the absence of acceptor. This equation is equivalent to equation (2) [20].

The values of $R_0$ can be defined [17-20] by the following expression:

$$R_0^6 = \frac{9000(\ln 10)k^2\phi_D}{128\pi^5 N n^4} \int_0^\alpha F_D(\lambda)\varepsilon_A(\lambda)\lambda^4 d\lambda \qquad (4)$$

The integral in the above equation, known as the spectral overlap integral $J(\lambda)$, consists of the normalized fluorescence spectrum of the donor, $F_D(\lambda)$, the extinction co-efficient spectrum of the acceptor, $\varepsilon_A(\lambda)$ (in M$^{-1}$cm$^{-1}$) and the wavelength $\lambda$ (in nm). Also, $\phi_D$ is the fluorescence quantum yield of the donor in the absence of acceptor and $n$ is the refractive index of the medium in the wavelength of the spectral overlap.

$$\text{Hence, } J(\lambda) = \int_0^\alpha F_D(\lambda)\varepsilon_A(\lambda)\lambda^4 d\lambda \quad (5)$$

The above definition of the $R_0$ can be rewritten in terms of $J(\lambda)$ with units $M^{-1}cm^{-1}nm^4$ as
$$R_0 = 0.2108(k^2 n^{-4}\phi_D J(\lambda)) \quad (6)$$

where $R_0$ is in units of $A^0$.

The disappearance or quenching of the donor fluorescence band in favour of acceptor fluorescence band is the evidence of efficient energy transfer [20]. In the present case the values of $J(\lambda)$, $R_0$ and $r$ are calculated using equations 5, 6 and 3 respectively.

## 4. Results and Discussions
### 4.1. Surface pressure-area isotherm

Figure 1 shows the surface pressure-area per molecule ($\pi - A$) isotherms of the monolayer of pure NK, RhB and NK-RhB mixtures (50:50 molar ratio) in water sub phase and in clay dispersion sub phase.

Surface pressure of RhB isotherm on pure water rises gradually up to 33.6 mN/m. At this point the floating monolayer of RhB is collapsed. These are in agreement with the reported literature [21]. NK isotherm is more steeper and surface pressure rises up to 58.9 mN/m before collapse pressure is reached. The isotherm of NK-RhB mixed monolayer lies in between the pure NK and RhB isotherm with respect to shape and collapse pressure. This indicates the miscibility or mixing of the two dyes in the mixed films [22, 23].

The shape and nature of the isotherms in the clay dispersion sub phase show some differences with respect to those in water subphase. The NK and NK-RhB mixed isotherms in presence of clay show distinct phases before collapse pressures are reached. Whereas RhB isotherm show only one phase ie. liquid-expanded phase. In this context it is important to mention that electrostatic interaction between the cationic dyes and clays are occurred at the air-water interface. This electrostatic interaction can also be described as an ion exchange reaction, as the charge compensating cations of the smectites are exchanged by the positively charged dye molecules [24]. Thus the clay particles come at the air-water interface and incorporated into Langmuir monolayer. Consequently it affects the monolayer characteristics.

### 4.2. AFM observation of monolayer LB film

To confirm the incorporation of clay particles and to have idea about the structure of the monolayer film, LB monolayer was studied by atomic force microscope (AFM). Figures 2a and 2b show the AFM images of LB monolayers of RhB in absence and presence of clay particle respectively, transferred onto smooth silicon substrates at a surface pressure of 15 mN/m along with the line analysis spectrum. AFM image of RhB monolayer LB film without clay shows a smooth surface indicating the uniform deposition of RhB without any aggregates. Since dimension of the individual dye molecules are beyond the scope of resolution, hence its not possible to distinguish individual RhB molecule. In RhB-clay hybrid film nano order clay particles are clearly visible and distinguishable. This is a clear indication that nano – clay platelets are incorporated into LB films. The surface coverage is almost 80%.The line analysis shows that the height of laponite domain in the films is in between 1.5 to 2.5 nm. But few clay domains with height 3.5 nm are also observed in the film. This indicates that laponite particles are partly overlapped in the films and they don't form smooth continuous layer.

### 4.3. Spectroscopic investigation

Figure 3a and b show the absorption and fluorescence spectra of pure NK, RhB and NK-RhB (50:50 molar ratio) mixture in chloroform solution ($10^{-6}$M). The pure NK absorption spectrum possess prominent intense 0-0 band at 430 nm along with a weak indistinguishable hump at 409 nm, which is assigned to be due to the 0-1 vibronic transition. The 430 nm 0-0 band is attributed to the NK monomer, whereas, 409 nm 0-1 band may be due to the contribution of NK H-dimer. On the other hand pure RhB absorption spectrum possesses prominent 0-0 band at 557 nm and weak band at 519 nm due to the vibronic transition. These two bands may be due to the trace of RhB monomer and dimer respectively. The absorption spectrum of the mixed solution contains all bands of the individual dyes. The shape and position of the bands are almost identical to their corresponding individual counterparts except a decrease in intensity. The decrease in intensity is due to the decrease in molar ratio of individual dyes in mixed solution. This indicates that there is no significant chemical and physical interaction among the dye molecules in the chloroform solution and mainly they remains as individual isolated components.

The fluorescence spectra of pure and mixed dyes are shown in figure 3b. The fluorescence spectra were obtained after excitation at 430 nm. The excitation wavelength was selected appropriately to excite the NK molecules directly and to avoid or minimize the direct excitation of the RhB molecules. The NK fluorescence spectra possess a broad prominent band at about 483 nm, whereas, less intense weak emission band at about 575 nm is obtained for pure RhB, which is almost a fraction of 50 in intensity if the excitation was at RhB emission maximum (at about 525 nm, figure not shown here). Our later studies also show that the fluorescence from RhB in clay dispersions as well as in LB films is almost negligible with excitation at 430 nm. Under these circumstances we can say that significant emission from RhB may be possible only after excitation via energy transfer from the NK molecules.

The NK-RhB mixture solution fluorescence spectrum is composed of the bands those for the individual components with an overall decrease in intensity with respect to their pure counterparts. This indicates that the intensity of the fluorescence band maxima in chloroform solution depends on the molar ratios of each dye and there is no energy transfer among the dyes in chloroform solution.

An important criteria for efficient intermolecular FRET is a good overlap between the emission spectrum of the donor (here NK) and the absorption spectrum of the acceptor (here RhB) [1,2]. Inset of figure 3b shows the normalized fluorescence spectrum of NK together with the normalized absorption spectrum of the RhB in chloroform solution. A closer look to the figure show that energy transfer from NK to RhB can occur because of the sufficient spectral overlap of the NK fluorescence and RhB absorption bands. In order to verify this possibility we have measured the fluorescence spectra of pure NK, RhB and NK-RhB mixture in chloroform solution, clay dispersion and in LB films in the absence and presence of clay particles.

Figure 4 shows the fluorescence spectra of dyes in clay dispersion. The dye loading was 10% of the Cation Exchange Capacity (CEC) of the clay. The trends in the fluorescence spectra for the individual dyes in the laponite dispersion are almost analogous to those in chloroform solution. The only changes are related to the slight shift of the bands. NK fluorescence band is blue shifted to 472 nm where as RhB fluorescence band is red shifted to 582 nm. One can also observe significant increase in NK fluorescence intensity in clay dispersion. Blue shifting and increase in fluorescence intensity for other dyes in constrained media have been reported [25,26] and explained to be as the conformational change of the monomer is responsible for the redistribution of energy levels. But our later studies of fluorescence spectra of dyes in LB films (figure 5a and b) show the presence of strong H-band. Therefore, we assume that here in clay dispersion the broad band at 472 nm may be an overlapping of both the monomer and H-band.

Although the theory predicts no fluorescence from H-aggregate but weak fluorescence from H-aggregates of various thiacyanine dyes has already been reported and attributed to the imperfect stacking of the dye molecules in the aggregates [27-29]. However, the most interesting observation of strong fluorescence with high fluorescence quantum yield from the H-aggregates of thiacyanine derivatives has been reported by C. Peyratont et. al [30] and explained due to an inclined molecular arrangement of the dye in the aggregates. So in the present case the dye NK may form aggregates in such a way that the transition dipole moments are inclined to each other by a small angle rather parallel.

The decrease in fluorescence intensity along with red shift of RhB absorption band in clay suspensions was reported [31] and attributed to the formation of dimers and consequent transition of excited energy from monomer to dimer.

It is interesting to mention in this context that absorption and intercalation of rhodamine 3B in various clays were studied by Ló´pez Arbeloa and co-workers [32-33, 35] and found red shifts of the monomer absorption maximum by 7, 12, and 12.5 nm on montmorillonite, 10, 17, and 23 nm on Hectorite, and 10 nm on Laponite. The smaller shifts in montmorillonite and Hectorite were attributed to monomer absorption on the external clay surface and the intermediate shifts were attributed to monomer adsorption in the interlamellar regions, and the larger shifts were attributed to dimers or higher order aggregates. Grauer et al. [34] found red shift of about 10 nm for rhodamine 6G in Laponite surface. Tapia Esté´vez and co-workers reported a red shift of the monomer absorption of rhodamine 6G by 11 and 24 nm in Laponite [36,37]. Whereas, Sasai et al. [38] found no wavelength shift when they studied the intercalation of rhodamine 6G in a fluor-taeniolate.

The adsorption of the dye in clay systems has generally been attributed to ion exchange [32-36], although Yariv et. al. has determined that the dye molecules adsorb in excess of their cation exchange capacity [39]. The red shifts have been attributed to a change in the polarity of the environment of the dye as it goes from a solution phase to the more polar solid clay phase [33-34, 36, 40].

The main quantitative changes in the fluorescence spectra occur for the reaction system containing the mixture of NK and RhB in the clay dispersion. Here the NK emission band almost diminishes in favour of RhB emission band. RhB emission band become significant. In this case the NK did not emit light, but the energy was transferred form NK (donor) to RhB (acceptor). This transferred energy excites more electrons followed by light emission from the RhB, which added to the original emission of RhB. As a result the intensity gets sensitized. Here the energy transfer occurs only on the clay dispersions and not in chloroform solution. It should be mentioned in this context that fluorescence resonance energy transfer (FRET) process is very sensitive to distances between the fluorophore (donor-acceptor) and occur only when the distance is of the order of 1-10 nm [1,2]. Therefore, here clay particles play an important role in determining the concentration of the dyes on their surfaces or to make possible close interaction between energy donor and acceptor component in contrast to inactive systems based on homogeneous solutions.

Figure 5a and b show the fluorescence spectra of dye system in LB films in absence and presence of clay particles respectively. Here also NK shows strong fluorescence. In absence of clay NK fluorescence is dominated by H-band with peak at 463 nm and the monomer band at 483 nm becomes almost indistinguishable. Whereas in presence of clay NK fluorescence contains overlap of both H-band and monomer band with in 460-490 nm region. For RhB the trend is very similar to clay dispersion and show a very weak fluorescence with peak at around 600 nm and 593 nm in absence and presence of clay particles respectively. Interestingly for mixed dye system energy transfer is observed for both the system. The NK fluorescence almost disappeared and RhB fluorescence becomes very strong due to the occurrence of FRET.

In clay dispersions and LB films with or without clay, we have observed energy transfer between NK and RhB. Using the equations 6,7,4 and 3 the values of spectral overlap integral, $J(\lambda)$, Förster radius ($R_0$), the energy transfer efficiency (E) and the distance between the donor and acceptor (r) are calculated for different systems under investigation and listed table 1. We have determined the the values of $\phi_D = 0.1$ for NK using standard procedure. Also we use n = 1.59 for LB films and 1.39 for clay dispersions and $k^2 = 0.5$ [41]. These values along with the spectroscopic data are used to quantify the FRET process.

Among the three systems we have studied, we found that the $R_0$ value in clay dispersion is largest, whereas, in films the $R_0$ values are close to each other. We calculate $J(\lambda)$ from the donor emission and acceptor absorption spectra. The higher value of $J(\lambda)$ in case of clay dispersion may be responsible for the larger value of $R_0$. The FRET efficiency is maximum in LB films with clay and minimum for LB films without clay. This may be due to the fact that the intermolecular distance between RhB and NK decreases when the dyes are adsorbed on clay surfaces and then transferred onto solid substrates. A schematic diagram of the dyes and clays in LB films are presented in figure 6.

Although we have observed the trace of FRET in between NK and RhB but our next step is to find best suitable conditions for efficient energy transfer. For this detail investigation on the systems is going on in our laboratory. For efficient energy transfer it is important to have monomer of the both donor and acceptor molecule. Otherwise it might be possible that the transfer of energy from monomer of donor to aggregates may contribute to the quenching or decrease of donor fluorescence.

## 5. Conclusion

In this communication we reported two dyes NK and RhB in Langmuir and Langmuir-Blodgett (LB) films with or with out a synthetic clay laponite. Distinct changes in the isotherms in water and clay dispersion sub phases indicate the formation of clay-dye hybrid monolayer. AFM image gives the compelling visual evidence that the laponite particles are incorporated in hybrid films. Spectroscopic characteristics reveal the nature of organization and aggregates of dyes in solution and in LB films. Trace of FRET was observed in clay dispersion and LB films. Efficiency of energy transfer was maximum in LB films with clay and minimum in LB films without clay.

**Acknowledgements**

SAH acknowledges the Fund for Scientific Research – Flanders for a postdoctoral fellowship. This research is funded by IAP- (Inter-University Attraction Pole) and by CECAT (Centre of Excellence – K. U. Leuven). DB and SAH are grateful to DST and CSIR, Government of India for providing financial assistance through DST Project No: SR/S2/LOP-19/07 and CSIR Project Ref. No. 03(1080)/06/EMR-II.

**Figure captions:**

Figure 1. Pressure-area isotherm at the air-water interface (a) with out clay and (b) with clay. 1 = 100%RhB, 2 = (RhB (50%) + NK (50%)), 3 = 100% NK. Inset: molecular structure of RhB and NK.

Figure 2a. AFM image of monolayer RhB without clay onto silicon substrate along with the line analysis spectrum.

Figure 2b. AFM image of monolayer RhB with clay onto silicon substrate along with the line analysis spectrum.

Figure 3. (a) UV-Vis absorption and (b) steady state fluorescence spectra of RhB and NK in chloroform solution. 1 = 100%RhB, 2 = (RhB (50%) + NK (50%)), 3 = 100% NK. Inset: molecular structure of RhB and NK. Inset: normalized fluorescence spectra of NK and UV-Vis absorption spectrum of RhB.

Figure 4. Fluorescence spectra of RhB and NK in clay suspension (dye loading is 10% of CEC of laponite). 1 = 100%RhB, 2 = (RhB (50%) + NK (50%)), 3 = 100% NK. Inset shows the enlarged fluorescence spectra of 100%RhB.

Figure 5. Fluorescence spectra of RhB and NK in (a) LB films without clay and (b) LB films with clay. 1 = 100%RhB, 2 = (RhB (50%) + NK (50%)), 3 = 100% NK. Inset shows the enlarged fluorescence spectra.

Figure 6. Schematic diagram of RhB, NK and clay in LB films.

Table 1. The values of spectral overlap integral, $J(\lambda)$, Förster radius ($R_0$), the energy transfer efficiency (E) and the distance between the donor and acceptor (r). All these values are calculated based on spectral characteristics.

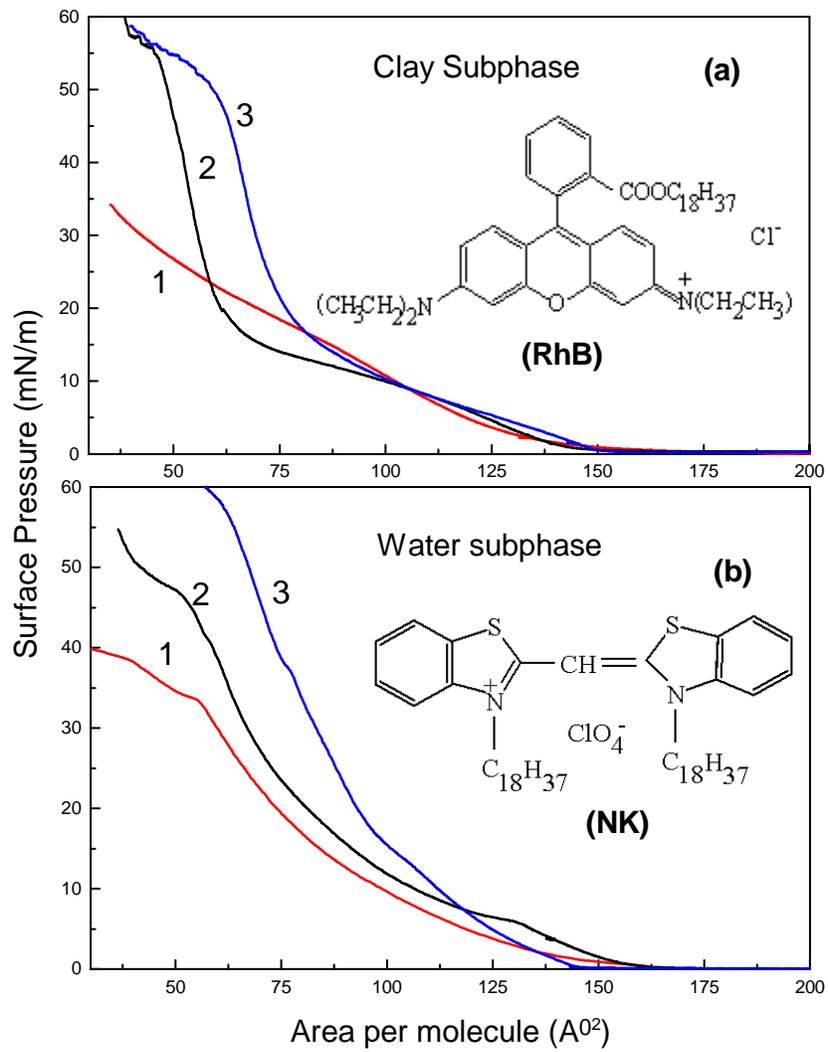

Figure 1 S. A. Hussain et. al.

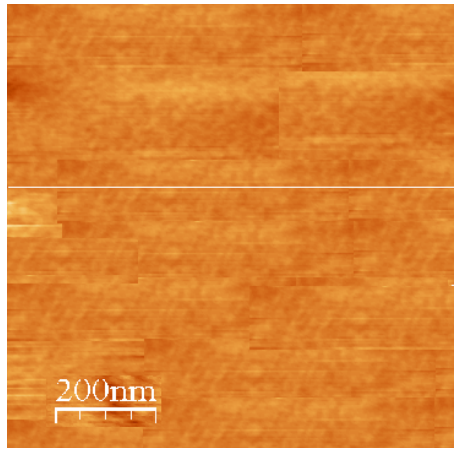
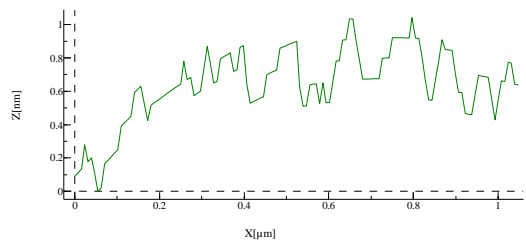

2a: S. A. Hussain et. al.

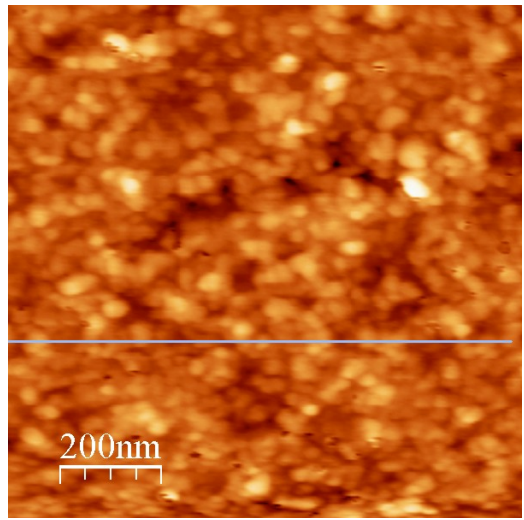
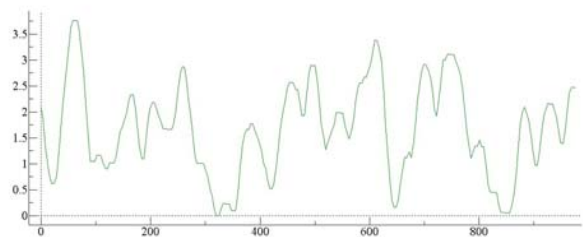

Figure 2b S. A. Hussain et. al.

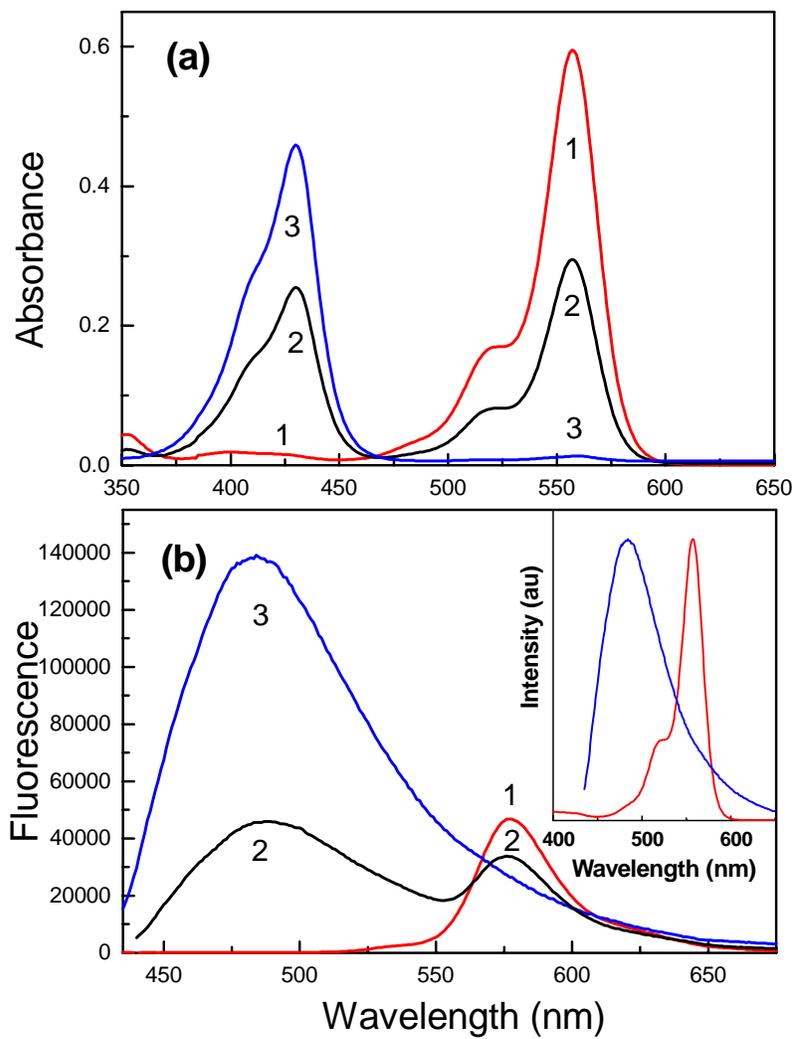

Figure 3 S. A. Hussain et. al.

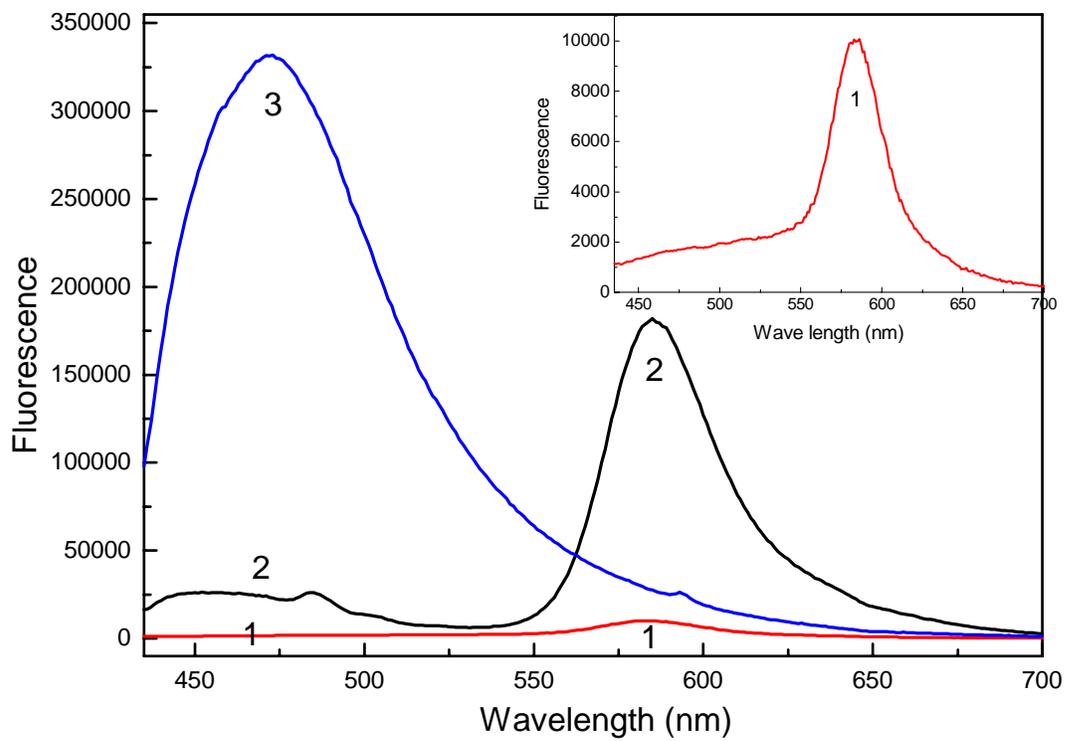

Figure 4 S. A. Hussain et. al.

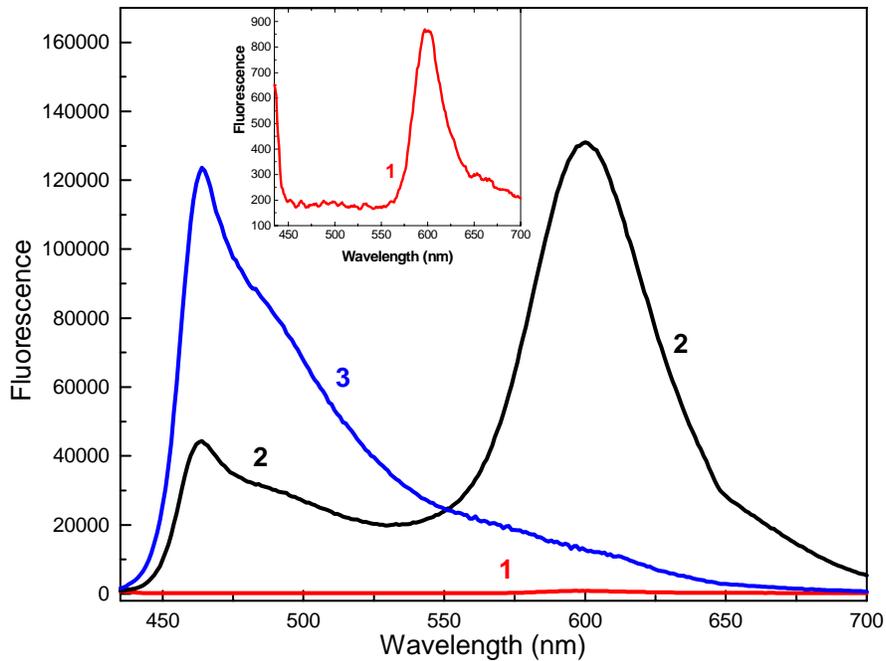

Figure 5a S. A. Hussain et. al.

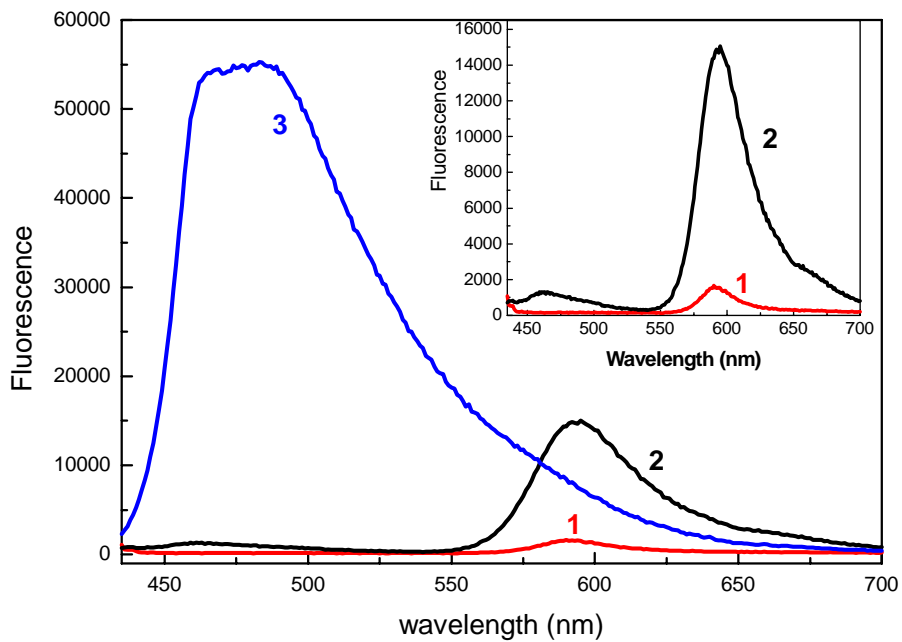

Figure 5b S. A. Hussain et. al.

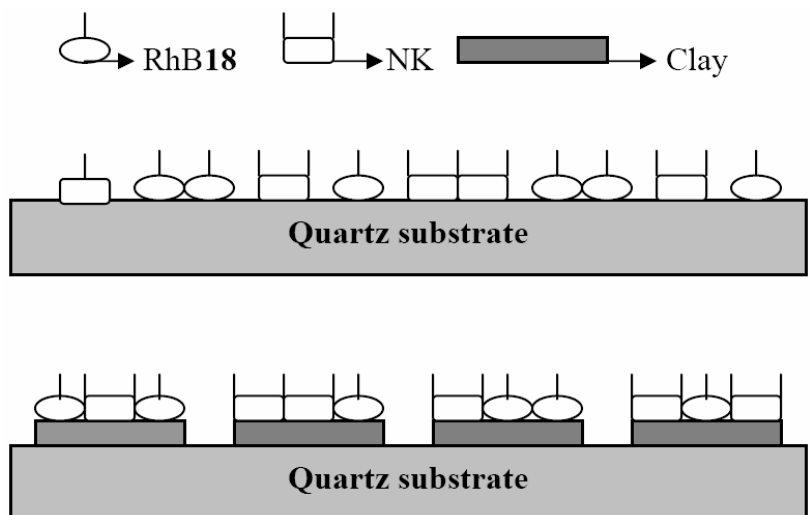

Figure 6 S. A. Hussain et. al.

|  | $J(\lambda)$ (M$^{-1}$cm$^{-1}$nm$^4$) | $R_0$ (A$^0$) | r (A$^0$) | E (%) |
|---|---|---|---|---|
| Clay suspension | $7.23 \times 10^{13}$ | 22.56 | 14.99 | 92.3 |
| LB film with clay | $1.84 \times 10^{13}$ | 15.54 | 8.36 | 97.7 |
| LB film without clay | $1.29 \times 10^{13}$ | 14.69 | 13.4 | 64.2 |

Table 1: S. A. Hussain et. al.